\documentclass{PoS}

\def\pt{$p_\mathrm{T}\,\,\,$}
\def\Dz{$D^0\,\,\,$}

\title{Production of open charm and beauty states in pPb collisions with LHCb}

\ShortTitle{LHCb open charm and beauty production in pPb data}

\author{\speaker{Yanxi ZHANG}\thanks{The corresponding author acknowledges support from the European Research Council
(ERC) through the project EXPLORINGMATTER, funded by the ERC through a ERC-Consolidator-Grant, until 31 August 2018.
}\\
        European Organisation for Nuclear Research (CERN)\\
        E-mail: \email{yanxi.zhang@cern.ch}}

\abstract{
    These proceedings summarize the LHCb measurements of charm- and beauty-hadron production in pPb collisions. The
    studies are made down to very low-\pt of the observed heavy-flavor hadrons using fully reconstructed decays. Nuclear
    matter effects are quantified via nuclear modification factors and forward-backward production ratios. A strong
    suppression is observed at positive rapidity (proton beam direction), while a modest or no suppression is seen for
    the backward rapidity (lead beam direction). The nuclear parton distributions of the lead nucleus is constrained
    down to Bjorken-$x\sim10^{-5}$, assuming it is the only nuclear effect for open heavy-flavor production. These
    data provide important inputs to understand the Quark-Gluon Plasma formed in heavy-nucleus collisions.
          }

\FullConference{International Conference on Hard and Electromagnetic Probes of High-Energy Nuclear Collisions\\
		30 September - 5 October 2018\\
		Aix-Les-Bains, Savoie, France}

\begin{document}

\section{Introduction}
Heavy-flavor quarks (charm and beauty) are unique probes of nuclear matter. Thanks to their heavy masses, they are produced
at the early times of the collisions,
and experience the whole evolution of the nuclear medium before hadronisation. As a result, their kinematics and
hadroniation contain information on the properties of the medium. The heavy-quark masses provide also a hard scale 
that allows theoretical predictions of their production based on perturbative QCD approach.
Studies of heavy-quark production in pPb collisions help to quantify cold nuclear matter effects (CNM), referring to those
affecting particle production that are not related to the presence of a deconfined medium, i.e. the Quark-Gluon Plasma. 
The most extensively discussed CNM effects include
the modification of collinear parton distribution functions described in the framework of nuclear PDF (nPDF)~\cite{nPDF} or
color-glass condensate (CGC)~\cite{CGC}. 
One way to quantity the CNM effects is to measure the nuclear modification factor, $R_{p\mathrm{A}}$, defined as the
ratio of the cross-sections in p-nucleus to pp collisions, scaled by the nucleus atomic number, $A$. A deviation from unity
of this factor for heavy-flavor production would indicate the presence of nuclear matter effect.

%
These proceedings present the measurements of $D^0$~\cite{LHCbD0}, $\Lambda_c^+$~\cite{LHCbLc}, $B^+$, $B^0$ and $\Lambda_b^0$
production~\cite{LHCbHb} in pPb
collisions at $\sqrt{s_\mathrm{NN}}=5\,\,\mathrm{and}\,\,8.16$ TeV by the LHCb collaboration.\footnote{Charge conjugated
states are included throughout this manuscript.} 
The pPb data correspond to an integrated luminosity of about 2 nb${}^{-1}$ and about 30 nb${}^{-1}$ for the charm and beauty
measurements. The data were collected with final state particles in either proton (forward, positive $y$) or lead beam (backward, negative $y$)
direction pointing into the LHCb acceptance from the interaction region.
The LHCb experiment was designed for precision measurements of beauty and charm hadrons in pp collisions, with optimized vertexing, tracking and
particle identification systems, and flexible trigger strategy, and now becomes a general-purpose detector. The detailed description of LHCb detector and
operation performances can be found in Refs.~\cite{LHCbDet,LHCbPer}. 


\section{Prompt $D^0$ production in pPb data at $\sqrt{s_\mathrm{NN}}=5$ TeV}
\begin{figure}[!tpb]
    \centering
    \includegraphics[width=0.40\textwidth]{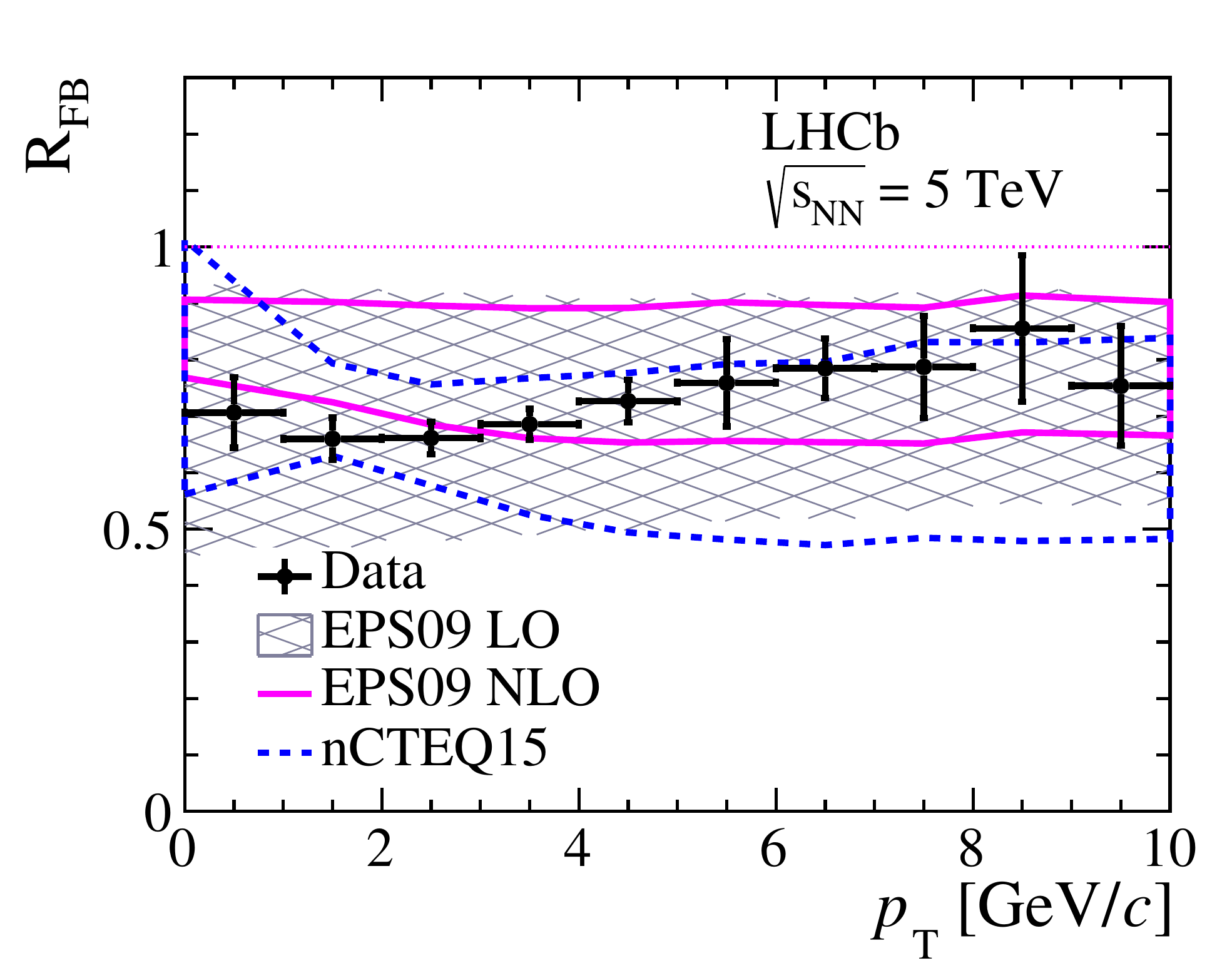}
    \includegraphics[width=0.40\textwidth]{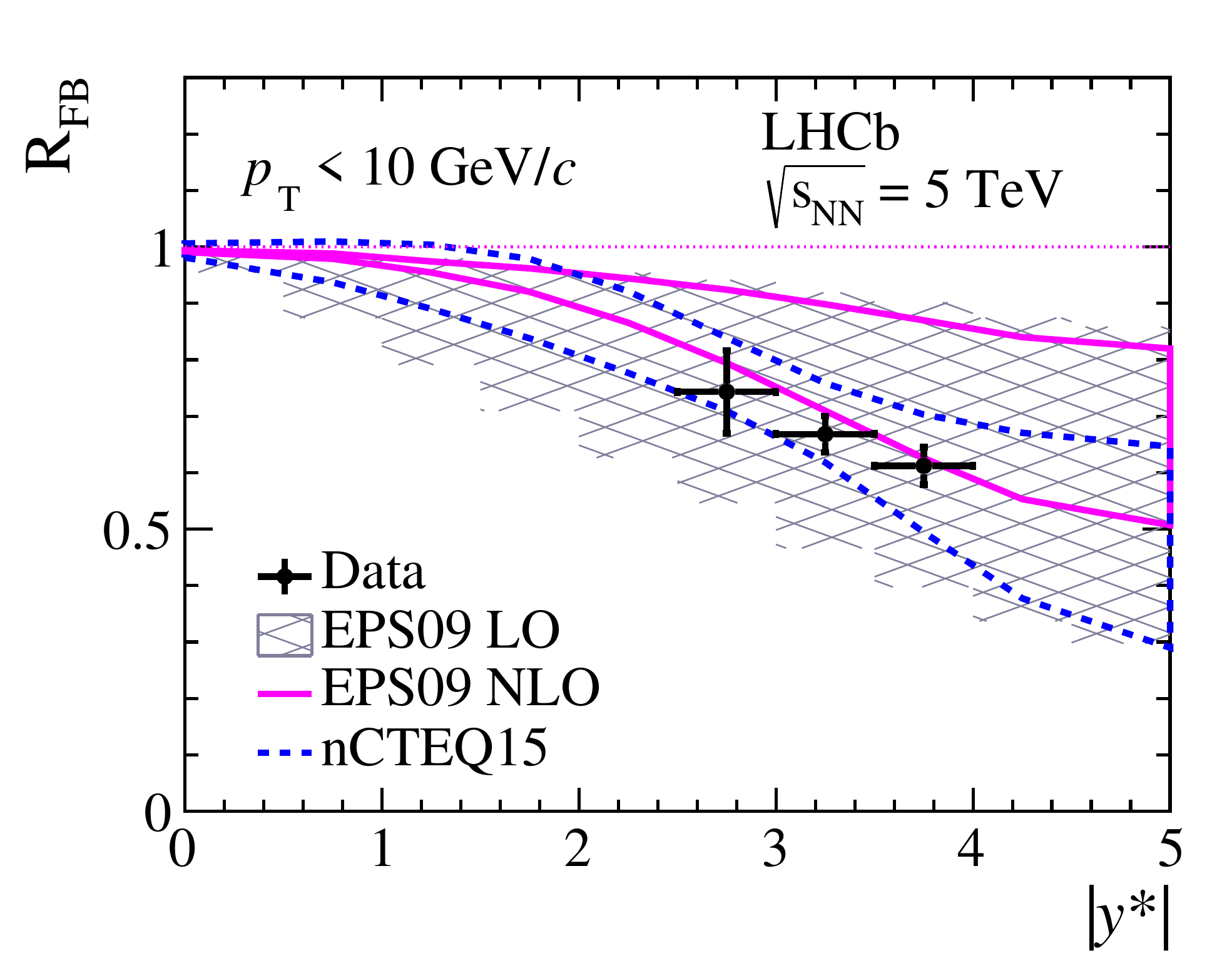}
    \includegraphics[width=0.40\textwidth]{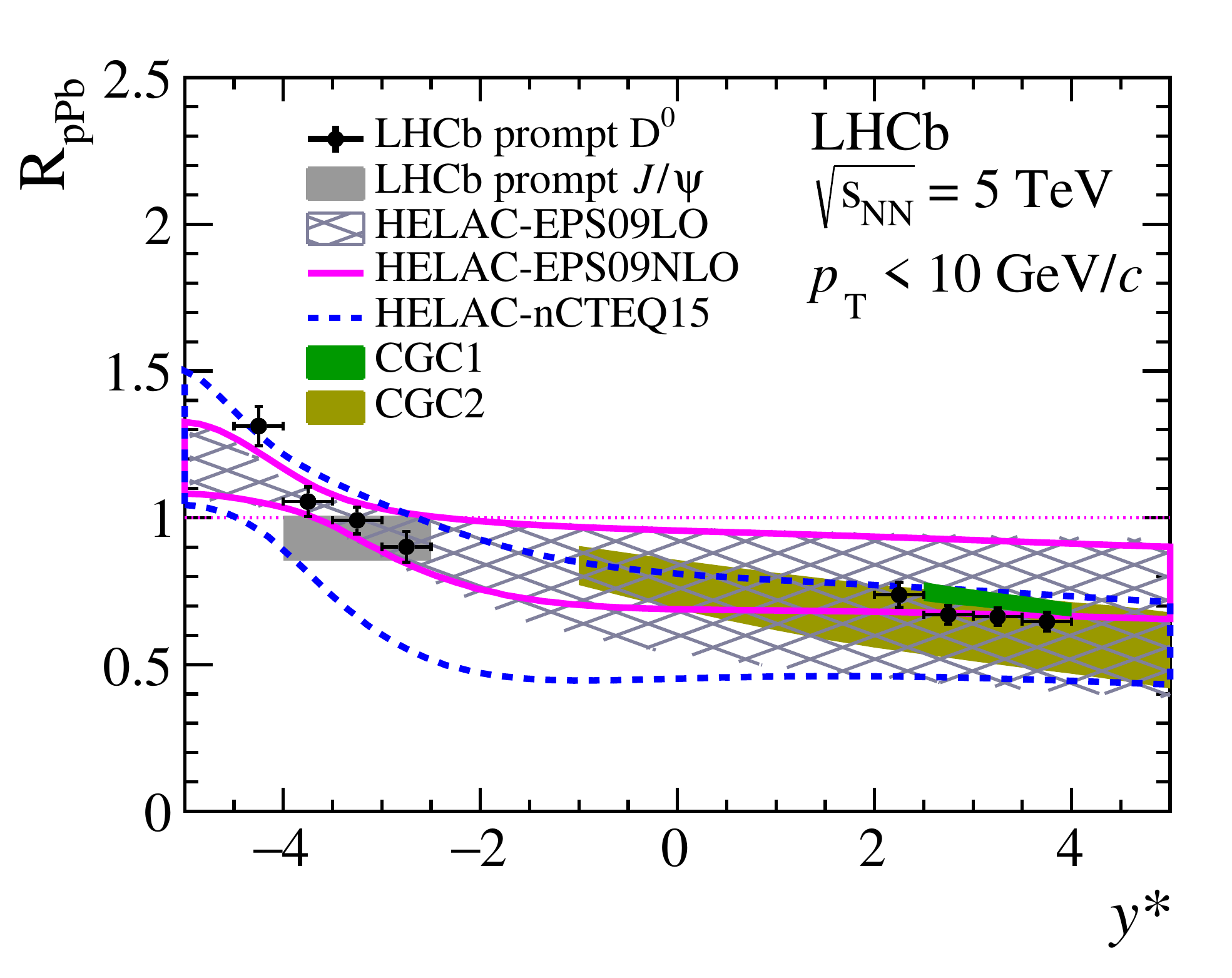}
    \includegraphics[width=0.40\textwidth]{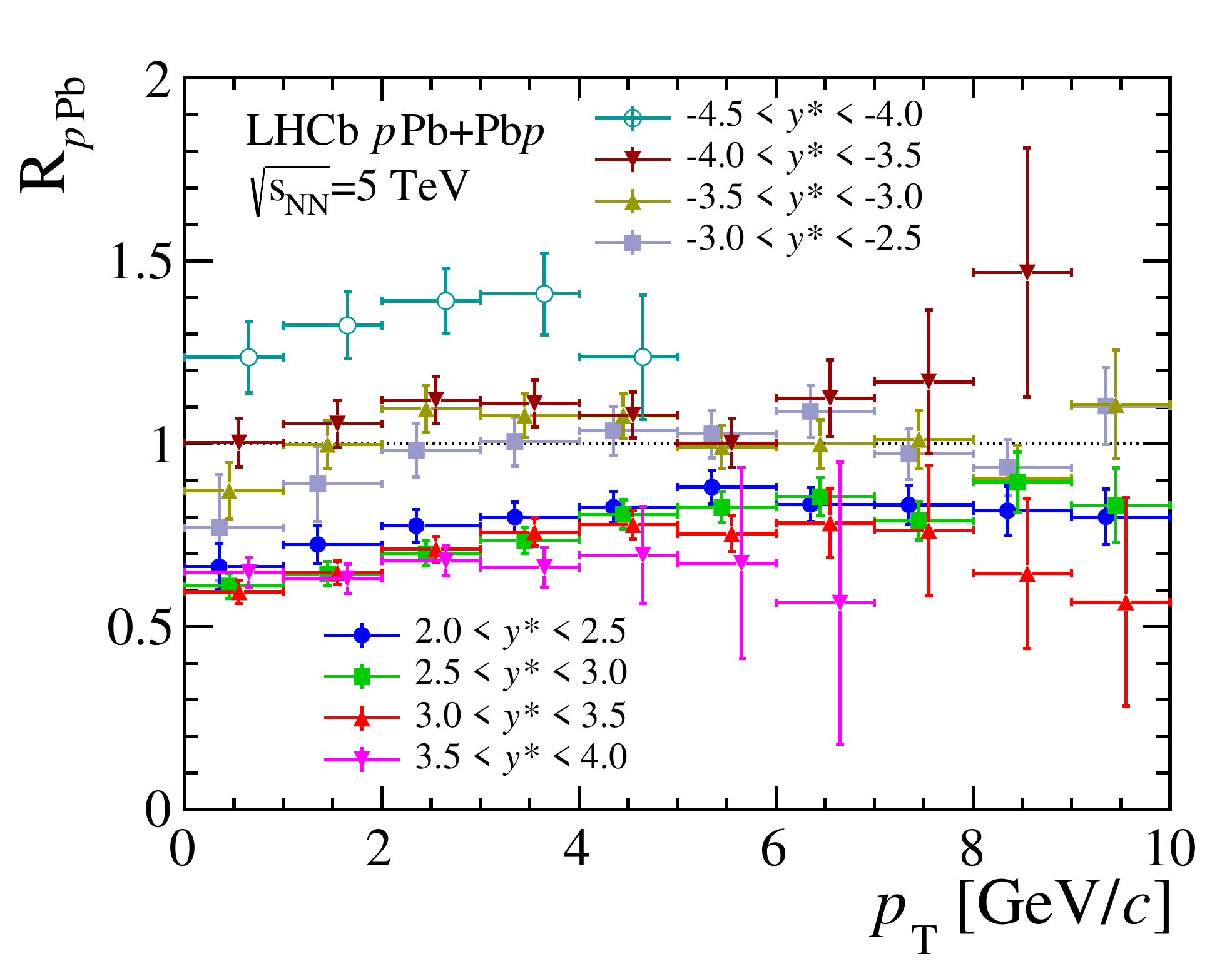}
    \caption{
        The \Dz meson $R_\mathrm{FB}$ as a function of (top left) \pt and (top right) $y$ of the hadron and
        $R_{p\mathrm{A}}$ as a function of (bottom left) rapidity and
        (bottom right) \pt for different rapidity bins. 
    }
    \label{figD0}
\end{figure}

The \Dz candidates are reconstructed through $D^0\to K^\pm\pi^\mp$ decays. Prompt \Dz and those feed-down from beauty
decays are separated by studying the \Dz impact parameter distribution. 
The forward-to-backward cross-section ratio, $R_\mathrm{FB}$,
quantifies the relative nuclear modification in the p- and the Pb-beam direction.  Displayed at the top of
Fig.~\ref{figD0}, the $R_\mathrm{FB}(D^0)$
indicates significant production asymmetry between p- and Pb-beam direction. The asymmetry increases at large 
absolute rapidity and reduces at high $p_\mathrm{T}$. If the effect is dominated by nPDF, it suggests a larger asymmetry of modifications of
gluon PDF at small and large-$x$. The measurements display reasonable agreements with calculations 
using different nPDF sets in the HELAC-Onia framework~\cite{HELAConia}. 
The measurement of $R_{p\mathrm{A}}$ for prompt \Dz as a function of \pt and $y$ is shown at the bottom of Fig.~\ref{figD0}. 
The $D^0$ $R_{p\mathrm{A}}$ as a function of rapidity shows a strong suppression ($\sim30\%$) at positive rapidity, while it is
compatible with no suppression at negative rapidity with a
hint of enhancement at extreme backward rapidity. The result suggests strong shadowing effect in the small-$x$ region and
hint of anti-shadowing in the appropriate large-$x$ area. Considering $R_{p\mathrm{A}}$ as a function of \pt
(Fig.~\ref{figD0} bottom right), 
the pattern is very similar for different rapidity bins, increasing from low-\pt to high-$p_\mathrm{T}$. 
The slope and magnitude of $R_{p\mathrm{A}}$ as a function of \pt vary only marginally across the rapidity bins for both pPb and Pbp data.
The $R_{p\mathrm{A}}$ for prompt \Dz mesons is also consistent with that of $J/\psi$ hadrons within current
experimental precision, and both agree with HELAC-Onia calculations using nPDF sets. At forward rapidity,
the \Dz $R_{p\mathrm{A}}$ is also described by CGC calculations~\cite{CGCD01, CGCD02} with chosen saturation scales
that model $J/\psi$ data.
The precision of the $D^0$  $R_{p\mathrm{A}}$ data has contributed to constraint significantly recent nPDF predictions~\cite{nPDFFits}.


\section{Prompt $\Lambda_c^+$ production in pPb data at $\sqrt{s_\mathrm{NN}}=5$ TeV}
\begin{figure}[!tpb]
    \centering
    \includegraphics[width=0.40\textwidth]{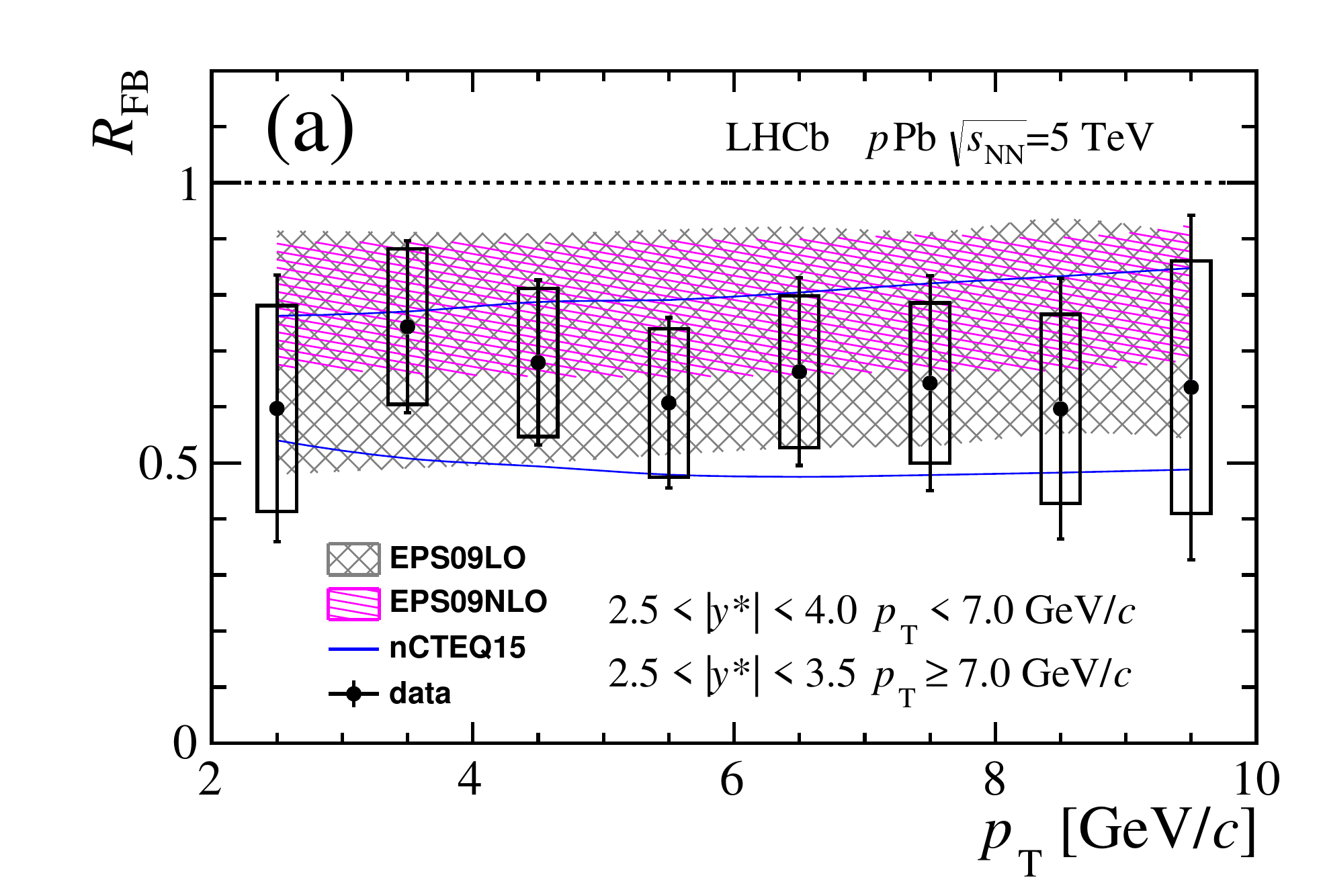}
    \includegraphics[width=0.40\textwidth]{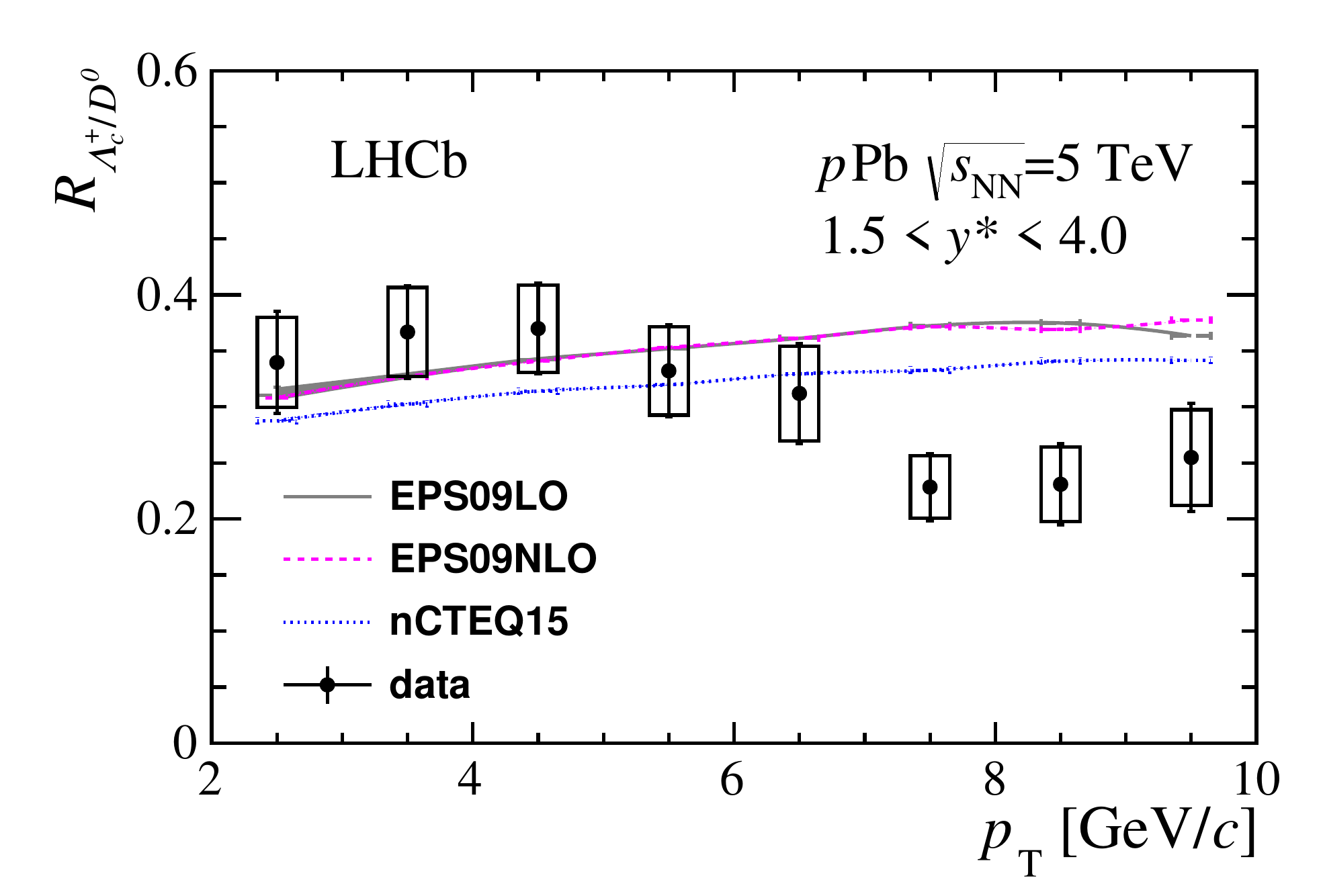}
    \caption{
        (Left) $R_\mathrm{FB}$ for prompt  $\Lambda_c^+$ production and (right) $\Lambda_c^+$ over $D^0$ production
        ratio in pPb data.
    }
    \label{figLcRlt}
\end{figure}

The $\Lambda_c^+$ baryon is reconstructed with $\Lambda_c^+\to pK^-\pi^+$ decays, resulting in
thousands of signals with little background contamination. The forward-backward ratio (left of Fig.~\ref{figLcRlt}),
being smaller than unity, indicates a stronger suppression
of $\Lambda_c^+$ production in the forward region compared with the negative rapidity, consistent with predictions using
nPDF sets. The charm baryon-to-meson ratio, $R_{\Lambda_c^+/D^0}$, which is sensitive to charm quark fragmentation, is also
studied. Shown
on the right of Fig.~\ref{figLcRlt}, the magnitude of $R_{\Lambda_c^+/D^0}$ is consistent with nPDF predictions, which
implies that cold nuclear
matter effect is almost identical for different charm-hadron species.
There might be a hint of discrepancy at high-$p_\mathrm{T}$, which will be followed up by studies using the larger sample of LHCb Run II
pPb data.
The ratio $R_{\Lambda_c^+/D^0}$ measured by LHCb, however, has tensions with the ALICE result that gives a value around
0.5 at low-\pt and mid rapidity~\cite{ALICELc}, more than 2-$\sigma$ deviations from LHCb. More detailed studies
are required to understand this puzzle using the larger pPb sample collected at $\sqrt{s_\mathrm{NN}}=8.16$ TeV.

\section{Beauty hadron production in pPb data at $\sqrt{s_\mathrm{NN}}=8.16$ TeV}
\begin{figure}[!tpb]
    \centering
    \includegraphics[width=0.40\textwidth]{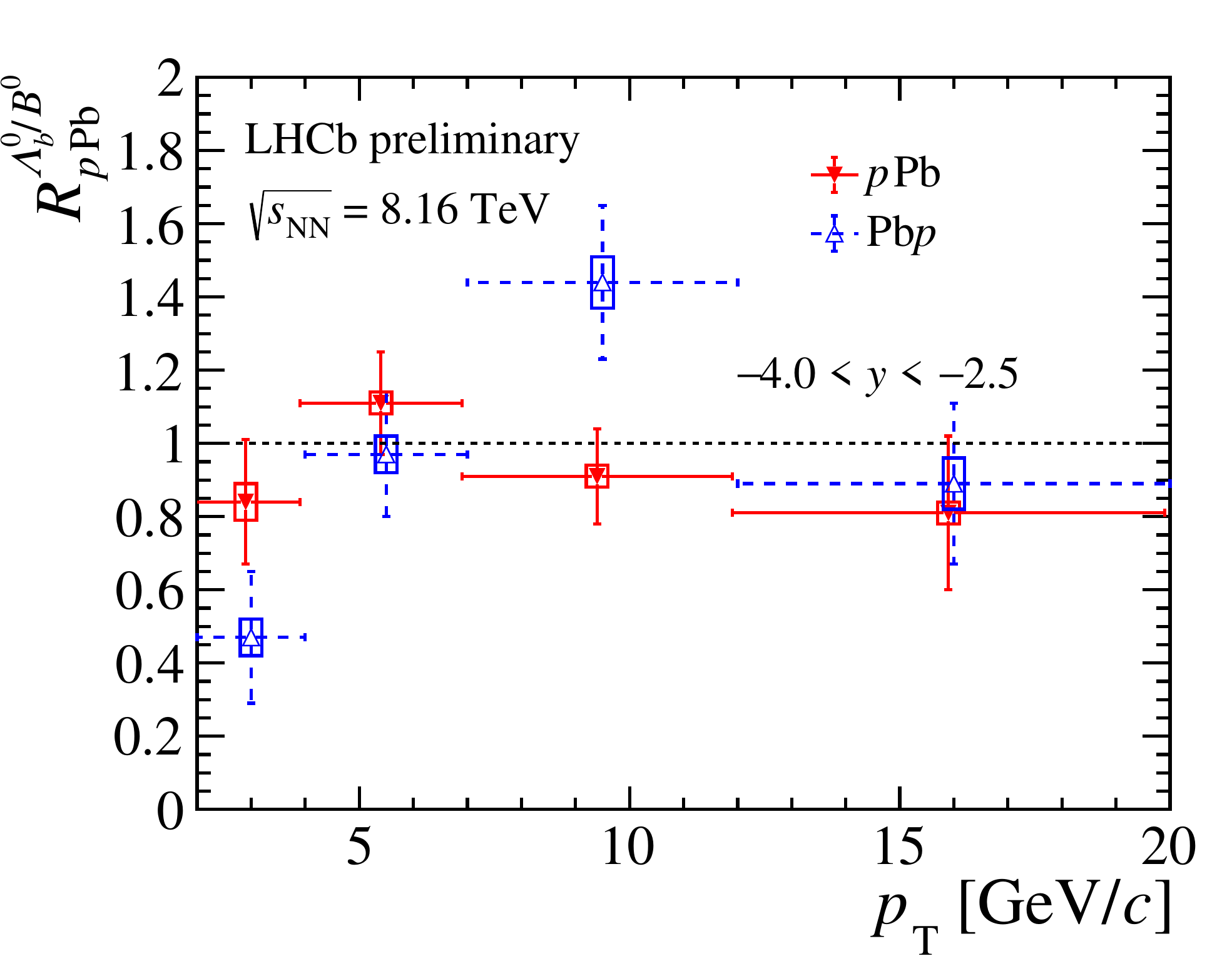}
    \includegraphics[width=0.40\textwidth]{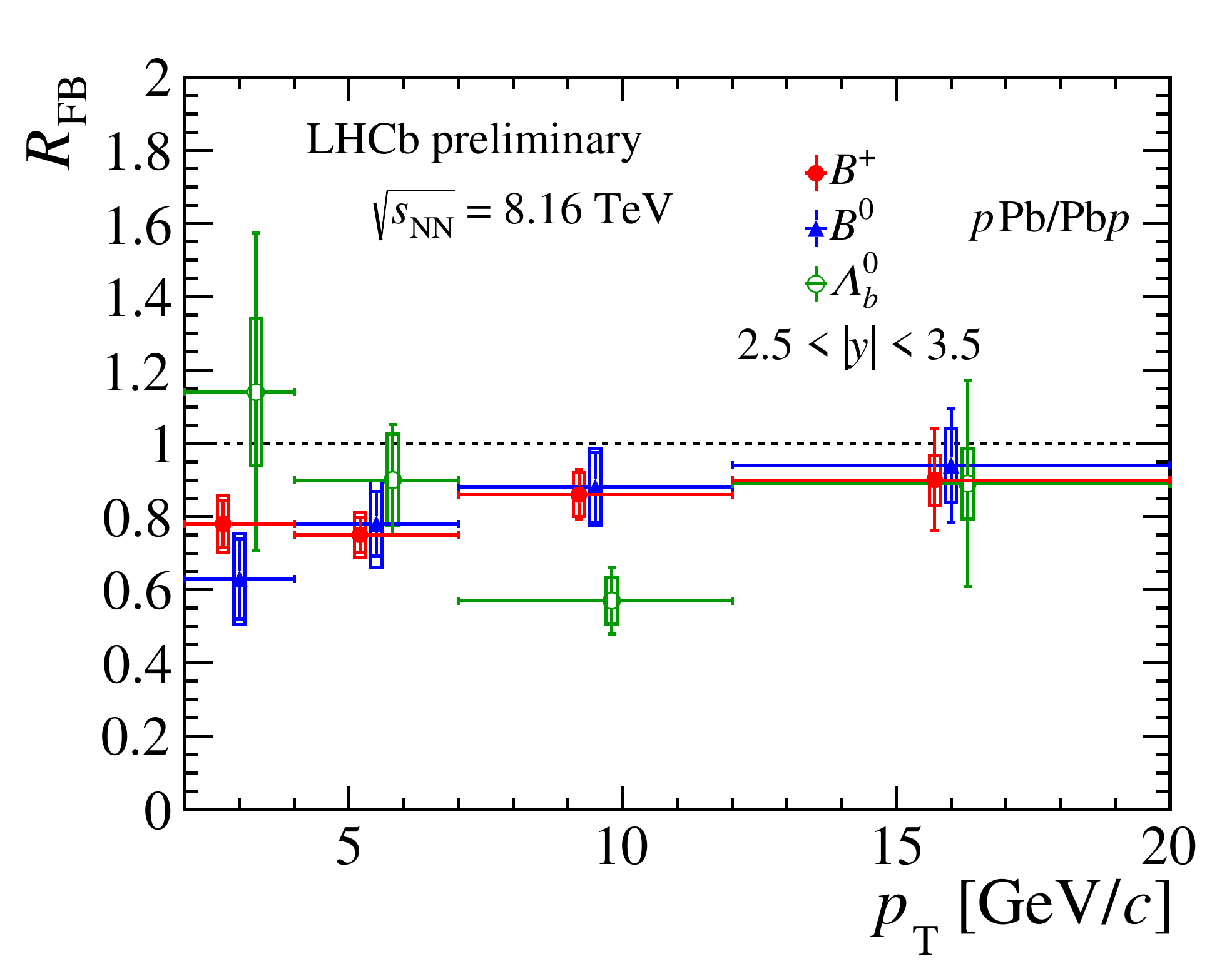}
    \includegraphics[width=0.40\textwidth]{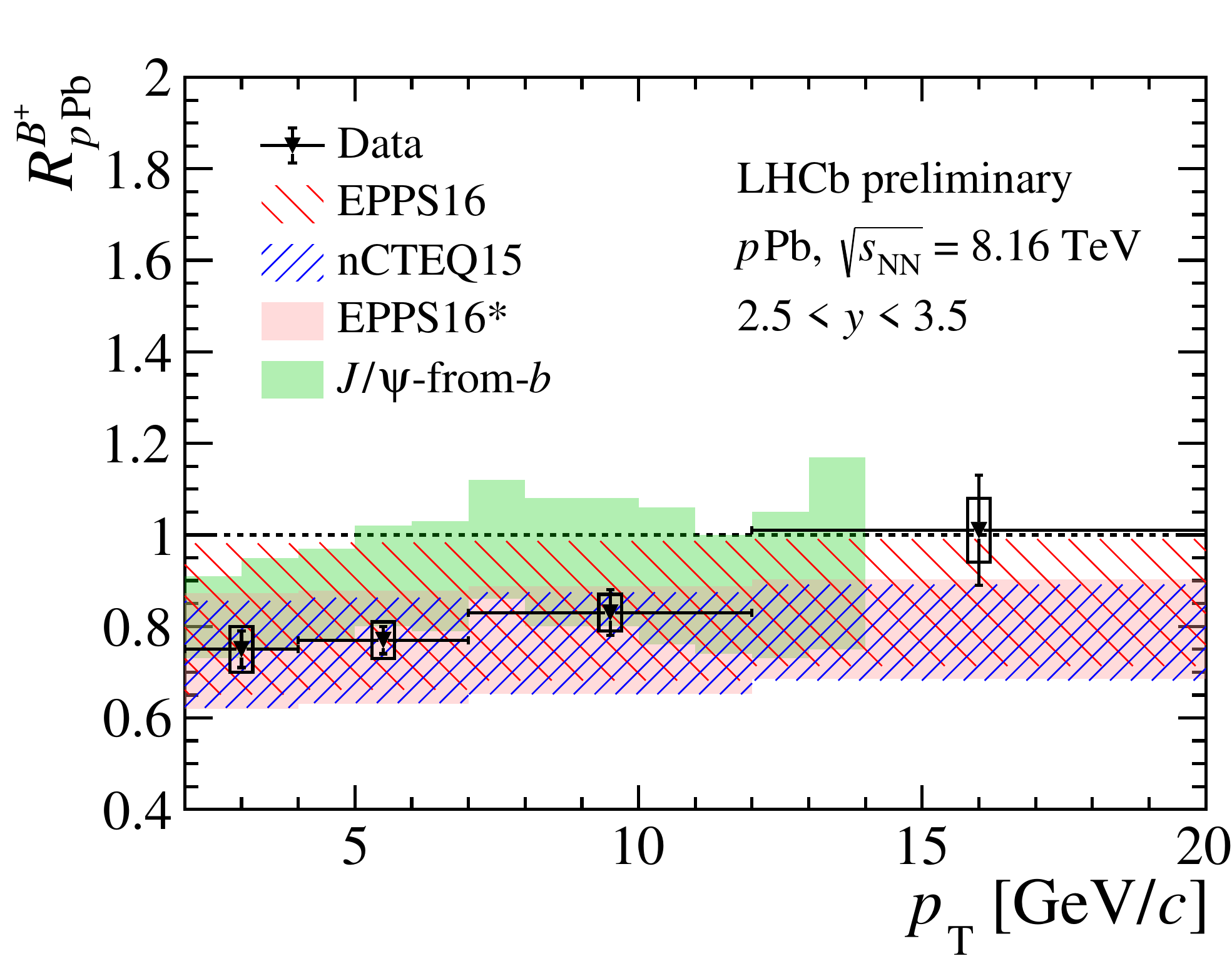}
    \includegraphics[width=0.40\textwidth]{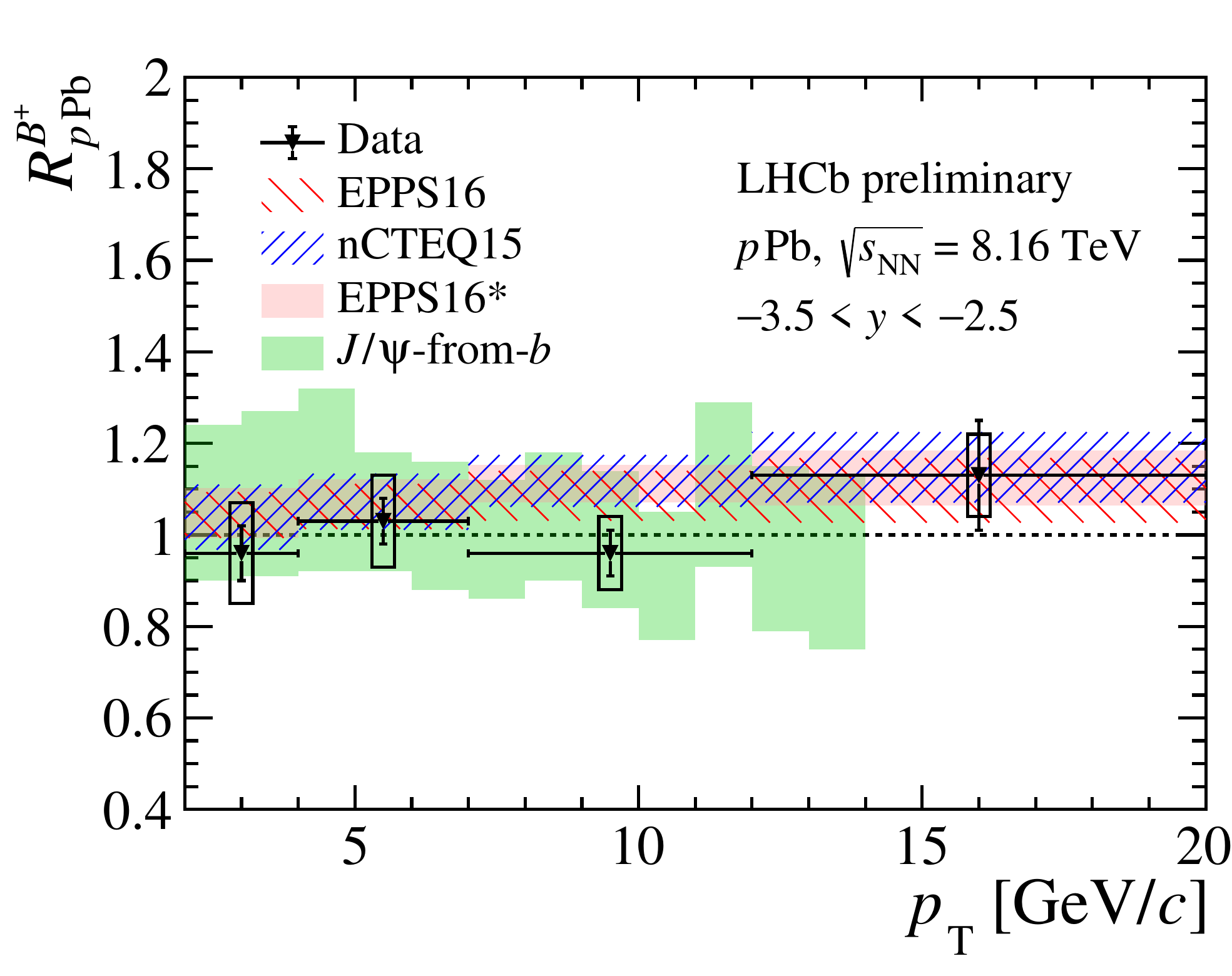}
    \caption{ (Top left) Beauty hadron $R_\mathrm{FB}$ and (top right) $\Lambda_b^0$ over $B^0$ production ratio.
     Nuclear modification of $B^+$ at (bottom left) negative and (bottom right) positive rapidity.}
    \label{figHb}
\end{figure}

The $B^+, B^0$ and $\Lambda_b^0$ production is studied using exclusive decays $B^+\to J/\psi K^+, \overline{D}^0\pi^+$,
$B^0\to D^-\pi^+$ and $\Lambda_b^0\to \Lambda_c^+\pi^-$. 
The $B^0/B^+$ cross-section ratio is found to be consistent with unity within one standard deviation, independent of \pt
or $y$, confirming the isospin symmetry in pPb collisions. The $\Lambda_b^0/B^0$ cross-section ratio is about 40\% averaging over the range
$2<p_\mathrm{T}<20$ GeV$/c$ and $2.5<|y|<3.5$. With a decreasing trend for increasing $p_\mathrm{T}$, the $\Lambda_b^0/B^0$
cross-section ratio reaches a value similar to LEP data~\cite{HFAG} at
$p_\mathrm{T}\sim 20$ GeV$/c$. The $\Lambda_b^0/B^0$  production ratio in pPb data over that in pp collisions, shown
on the top left of Fig.~\ref{figHb}, is compatible with unity in all \pt bins, which suggests a similar nuclear effect
between the beauty baryon and mesons. Displayed on the top right of Fig.~\ref{figHb} is the measurement of
$R_\mathrm{FB}$ for $B^+, B^0$ and $\Lambda_b^0$. Results are similar for the three beauty-hadron species. 
The $R_\mathrm{FB}$ for $B^+$, which has a better experimental precision, is smaller than unity and is in good agreement with nPDF calculations. 
The $R_{p\mathrm{Pb}}$ is also calculated for the $B^+$ hadron, depicted at the bottom of Fig.~\ref{figHb} as a function of
$p_\mathrm{T}$. The patten is similar to that of $D^0$. The result shows significant suppression
of $B^+$ production at low-\pt at positive rapidity, which almost vanishes going to high-$p_\mathrm{T}$. At backward
rapidity, the $R_{p\mathrm{Pb}}$ for the $B^+$ hadron is consistent with unity and demonstrates no obvious \pt dependence. The measurement agrees with
predictions involving nPDF sets, and specifically the nPDF set that is constrained by LHCb data of prompt $D^0$ production
in pPb collisions. The result also confirms previous LHCb measurement using $J/\psi$ from $b$-hadron decays. If
the modification of PDF in nucleus is the dominant nuclear effects, $R_{p\mathrm{A}}$ for different beauty-hadron
species is predicted to be almost identical, as is the case for charm hadrons. 
With the experimental uncertainty being smaller than those for theoretical calculations at forward rapidity, LHCb beauty-hadron
measurement provides further constraints to nPDF.

\section{Summary}
With the precise and diverse measurements of charm and beauty hadrons in pPb collisions, LHCb made important
contributions to the studies of heavy-ion physics.
Prompt $D^0$ cross-section is measured down to zero-$p_\mathrm{T}$, showing strong
suppression for production in proton beam direction. The cross-section ratio of $\Lambda_b^0$ over
$B^0$ is consistent with expectation from pp collisions with current precision, indicating similar nuclear effects for
beauty baryon and meson production in pPb collisions.  
More measurements involving charm and beauty hadron are under study using the new pPb data samples collected at $\sqrt{s_\mathrm{NN}}=8.16$ TeV.

\begin{thebibliography}{99}
\bibitem{nPDF}
    M. Hirai, S. Kumano, and T.-H. Nagai, 
        \emph{Phys. Rev. C} {\bf 76} (2007) 065207
        [{\tt arXiv:0709.3038}]
\bibitem{CGC}
    F.Gelis et al,
        \emph{Ann. Rev. Nucl. Part. Sci.} {\bf 60} (2010) 463
        [{\tt arXiv:1002.0333}]
\bibitem{LHCbD0}
   R. Aaij et al, LHCb collaboration, 
        \emph{JHEP} {\bf 1710} (2017) 090
     [{\tt arXiv:1708.02750}]
\bibitem{LHCbLc}
   R. Aaij et al, LHCb collaboration, 
        \emph{To be published in JHEP}
     [{\tt arXiv:1809.01404}]
\bibitem{LHCbHb}
   R. Aaij et al, LHCb collaboration, 
        \emph{LHCb-CONF-2018-004}
     [{\tt arXiv:LHCb-PAPER-2018-48}]

\bibitem{LHCbDet}
   A.A. Alves Jr. et al, LHCb collaboration,  
        \emph{JINST} {\bf 3} (20108) S08005
\bibitem{LHCbPer}
   R. Aaij et al, LHCb collaboration,
        \emph{Int. J. Mod. Phys. A} {\bf 30} (2015) 1530022
     [{\tt arXiv:1412.6352}]
\bibitem{HELAConia}
   H.-S. Shao, 
        \emph{Comput. Phys. Commun.} {\bf 198} (2016) 238
     [{\tt arXiv:1507.03435}]
\bibitem{CGCD01}
   B. Ducloue, T. Lappi, and H. Mantysaari,
    \emph{Phys. Rev. D} {\bf 91} (2015) 114005
     [{\tt arXiv:1503.02789}]
\bibitem{CGCD02}
   H. Fujii and K. Watanabe,
    \emph{JLAB-THY-17-2506 }
     [{\tt arXiv:1706.06728}]
\bibitem{nPDFFits}
   A. Kusina et al, 
    \emph{Phys. Rev. Lett.} {\bf 121} (2018) 052004
     [{\tt arXiv:1712.07024}]
\bibitem{ALICELc}
   S. Acharya et al, ALICE collaboration, 
    \emph{JHEP} {\bf 04} (2018) 108
     [{\tt arXiv:1712.09581}]
\bibitem{HFAG}
   Y. Amhis et al, HFLAV group, 
    \emph{Eur. Phys. J. C} {\bf 77} (2017) 895
     [{\tt arXiv:1612.07233}]
\end{thebibliography}
\end{document}